# A Community Authorization Service for Group Collaboration

Laura Pearlman
Information Sciences Institute
The University of Southern California
laura@isi.edu

Von Welch
Department of Computer Science
The University of Chicago
welch@mcs.anl.gov

Ian Foster
Mathematics and Computer Science
Division, Argonne National Laboratory
Department of Computer Science
The University of Chicago
foster@mcs.anl.gov

Carl Kesselman
Information Sciences Institute
The University of Southern California
carl@isi.edu

Steven Tuecke
Mathematics and Computer Science
Division, Argonne National Laboratory
tuecke@mcs.anl.gov

**Abstract**

*In "Grids" and "collaboratories," we find distributed communities of resource providers and resource consumers, within which often complex and dynamic policies govern who can use which resources for which purpose. We propose a new approach to the representation, maintenance, and enforcement of such policies that provides a scalable mechanism for specifying and enforcing these policies. Our approach allows resource providers to delegate some of the authority for maintaining fine-grained access control policies to communities, while still maintaining ultimate control over their resources. We also describe a prototype implementation of this approach and an application in a data management context.*

## 1 Introduction

The sharing and coordinated use of resources within large, dynamic, multi-institutional communities is fundamental to an increasing range of computer applications, ranging from scientific collaboratories to healthcare. This sharing may involve not only file exchange but also direct access to computers, software, data, and other resources, as is required by a range of collaborative problem-solving and resource-brokering strategies emerging in industry, science, and engineering.

This sharing is, necessarily, highly controlled, with resource providers and consumers defining clearly and carefully just what is shared, who is allowed to share, and the conditions under which sharing occurs.

A set of individuals and/or institutions defined by such sharing rules form what has been called a *virtual community* or *virtual organization* (VO) [1]. Infrastructures that support the creation and operation of VOs are often called Grids [2]. Specialized Grids focused on close and routine interactions between people, instruments and information in support of widely distributed scientific research projects are often called *collaboratories* [3].

A key problem associated with the formation and operation of distributed virtual communities is that of how to specify and enforce community policies. Consider the situation in which a multi-institutional project has received an allocation of time on a shared computational resource. With current technologies, each change in personnel at participating institutions requires that the project leader contact the resource owner to create an account and allocation for each new team member. Furthermore, as project policies change, the project leader will have to go back to the resource provider to adjust allocations, rights and priorities for the team members so that they are consistent with the current focus of the collaboration. This interaction places undue burdens on resource providers, who are, in effect, forced to implement the policy decisions of the consortium. Conversely, these interactions also place significant overhead on the administration of the consortium, as every policy alteration can require interactions with every resource provider with which the project has established a relationship.

Policy enforcement for VOs comprising multiple institutions and resource providers imposes unique challenges, as follows.

*Scalability*. The cost of administering a VO (e.g., adding or removing participants, changing community policy) should not increase with the number of resource providers participating in the VO. Resource administration overheads should also be bounded. As each VO represents a new policy, it is reasonable to require that the cost of administering a resource should be proportional to the number of VOs, and not their size or dynamics.

*Flexibility and expressibility*. Community policy can be idiosyncratic. It may apply to sets of resources (e.g., restricting what fraction of total storage capability available to the VO) and will vary over time. Enforcing these agreements and policies in a distributed fashion introduces difficult bookkeeping issues.

*Policy Hierarchy*. VOs may be hierarchical. For example, each institution within a collaboration may wish to define its own institutional policy. Each of these nested policies must be consistent: institutional policy must be consistent with the VO policy, which must in turn be consistent with the local policy defined by each resource.

We argue that the solution to these and related problems is to allow resource owners to grant access to blocks of resources to a community as a whole, and let the community itself manage fine-grained access control within that framework.

We have designed and implemented a *Community Authorization Service* (CAS) that provides this capability. A community runs a CAS server to keep track of its membership and fine-grained access control policies. A user wishing to access community resources contacts the CAS server, which delegates rights to the user based on the request and the user's role within the community. These rights are in the form of capabilities [4], which users can present at a resource to gain access on behalf of the community. The user effectively gets the intersection of the set of rights granted to the community by the resource provider and the set of rights defined by the capability granted to the user by the community.

The CAS architecture builds on public key authentication and delegation mechanisms provided by the Globus Toolkit [5] Grid Security Infrastructure (GSI) [6], a widely used set of authentication and authorization mechanisms that address single sign on, delegation, and credential mapping issues that arise in VO settings.

As compared to other authorization systems such as Akenti [7, 8] and Secure Virtual Enclaves [9], CAS provides mechanisms for distributing administration that are critical for solving the issues of scalability and flexibility. Neuman proposes but does not implement a similar idea [10].

To date, we have applied CAS to one application, file access control. However, we believe that the combination of CAS with restricted delegation and standard interfaces to policy engines has broad utility in providing a scalable policy method for communities, and we and others plan to apply it in many other contexts in the near future.

The rest of this article is as follows. In Section 2, we review GSI features essential to understanding our implementation of CAS. In Section 3 we discuss the CAS architecture in detail. In Section 4 we explain major extensions we made to the GSI mechanisms to support CAS. In Section 5 we discuss our current implementation. In Section 6 we present a case study of integration with a real world data access application. In Section 7 we discuss security considerations. In Section 8 we review related work. In Section 9 we discuss future directions.

## 2  Grid Security Infrastructure

The Globus Toolkit's Grid Security Infrastructure (GSI) [6] has emerged as an essential middleware component that has been integrated into many tools, including FTP and LDAP servers and clients and remote job submission tools. GSI shares with SSH and Kerberos the distinction of being a security solution that has been deployed at a number of Grids including the NASA Information Power Grid [11] and the NSF PACI Grids [12, 13]. Like these technologies, GSI consists of a set of standard interfaces and protocols and an implementation. We summarize here those GSI features needed to understand our CAS implementation: proxy credentials, delegation, and authorization.

### 2.1  Overview

The Grid Security Infrastructure (GSI) software is a set of libraries and tools that allow users and applications to access resources securely. GSI focuses primarily on authentication and message protection [14], defining single sign-on algorithms and protocols, cross-domain authentication protocols, and delegation mechanisms [15-18] for creating temporary credentials for users and for processes executing on a user's behalf.

GSI, building on earlier work described in [19], is based on Public Key Infrastructure (PKI) and uses authentication credentials composed of X.509 certificates and private keys. In brief, a GSI user generates a public-private key pair and obtains an X.509 certificate from a trusted entity called a Certificate Authority (CA). These credentials then form the basis for authenticating the user to resources on the Grid.

One change from the GSI model described in [19] is that GSI now uses temporary credentials called proxy credentials. Described in detail in the following subsection, proxy credentials allow GSI to support single sign-on by allowing users to access resources at multiple sites without repeated authentication and to delegate their rights to remote processes. This single sign-on capability is critical to advanced Grid applications in which a single interaction may involve the coordinated use of resources at many locations.

### 2.2 Proxy Credentials

In the Grid environment, a user may need to authenticate multiple times in a relatively short period of time in order to coordinate access to multiple resources. Requiring users to type their pass phrase repeatedly is undesirable both from a convenience and security standpoint, as each decryption of the private key provides another opportunity for it to be compromised.

GSI solves this problem with proxy credentials. A user creates a proxy credential by generating a new private-public key pair and then generating a new certificate signed using the private key from the user's long-term credential. This process essentially creates a short-term binding between the new key pair and the user's identity. To authenticate using a proxy credential, a user presents both the proxy certificate and the long-term certificate. The relying party then verifies that the long-term certificate is valid, that the long-term certificate's private key was used to sign the proxy certificate, and that the user can demonstrate proof of possession of the proxy certificate's private key. If these conditions (and some others regarding certificate format and lifetime) are met, then the authentication succeeds, and the user is considered to have the identity associated with the long-term certificate. GSI implements this authentication process via the TLS protocol in a GSS-API library.

In order to support CAS we have also introduced the ability for a proxy to carry policy information restricting its use. We call a proxy carrying such a policy a *restricted proxy*. We have standardized the format for these proxy credentials and have submitted it as an internet draft to the IETF PKIX working group [20].

### 2.3 Delegation

It is often important in distributed applications for a user's application to be able to act, unattended, on the user's behalf. For example, a user may submit a job to a remote site, and that job may in turn need to access some of the user's files stored on a mass storage system located at a third site.

GSI allows the user to delegate a proxy credential to a process on a remote host; in the example above, the user can delegate a credential to the remote job, which can then use that credential to authenticate to the mass storage system on behalf of the user. The relying party performs the same verification as described for proxy certificates in Section 2.2, verifying the signatures in the whole certificate chain.

### 2.4 Authorization

GSI supports the notion of local policy enforced locally. To achieve this, GSI provides mechanisms for translating a user's GSI identity (i.e., the distinguished name from the user's certificate) to a local identity (for example, a Kerberos principal, or a local Unix user account). Once translated, the local identity can be used to enforce local policy decisions, such as file access, disk quotas, and CPU limits.

The Community Authorization Service, described next, augments the existing local policy enforcement mechanisms provided via GSI. It enables community policy to be enforced on the basis of the user's GSI identity, which is constant across resources, rather than the local identity that will vary from resource to resource. It is because the CAS works with GSI identities that scalability with increased resource count is achieved.

## 3 Community Authorization

As we indicated in the introduction, the fundamental problem that we address in this work is the scalable representation and enforcement of access policy within distributed virtual communities. Such communities may comprise many members, each participating as resource provider and/or resource consumer. In such settings, expressing policies in terms of direct trust relationships between producers and consumers has the problems of scalability, flexibility, expressibility, and lack of policy hierarchy.

We address these problems by introducing a trusted third party, a *community authorization service* (CAS) server that is responsible for managing the policies that govern access to a community's resources.

The CAS server contains entries for CAs, users, servers and resources that comprise the community and groups that organize these entities. It also contains policy statements that specify *who* (which user or group) has the permission, *which* resource or resource group the permission is granted on, and *what* permission is granted. *What* permission is denoted by a *service type* and an *action*; the *action* describes the type of action (e.g., "read" or "execute program"), and the *service type* defines the namespace in which the action is defined. Different resource servers may recognize different service types, but all resource servers that recognize the same service type should share the same interpretation of that service type's actions.

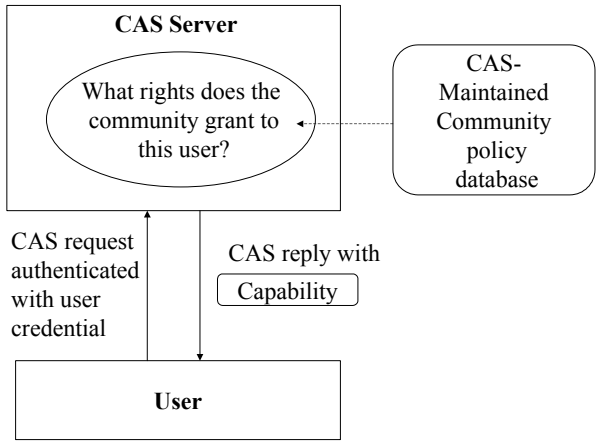

**Figure 1: In order to gain access to a CAS-managed community resource, a user must first acquire a capability from the CAS server.**

As illustrated in Figure 1, a member of a community may send the CAS server a request for a capability that will allow the user to perform a set of actions; if that request is consistent with the community's policy, the CAS server will delegate an appropriate capability back to the user. The user can then use that delegated credential to authenticate to a resource server and exercise the rights described by the capability. Of course, this authentication and exercise of rights is effective only if the resource provider has granted those rights to the community.

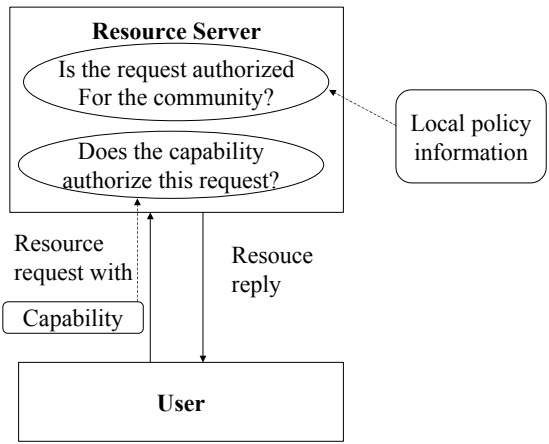

**Figure 2: A CAS capability is used to authenticate to the resource. This action can be repeated using the same capability until it expires.**

As shown in Figure 2, when the user presents the capability to a resource, the resource grants the user access to the local community resources based on local policy for the community (determined using the resource server's normal local access control mechanisms) and the community policy for the user (determined by examining the policy statements carried in the capability). In other words, the resource server will permit a request authenticated with a capability if the resource server's local policy authorizes the request for the grantor of the capability, and the capability itself authorizes the request for the bearer.

This structure addresses the scalability problem by reducing the necessary trust relationships from $CxP$ to $C+P$: each consumer needs to be known, and trusted, by the CAS server, but not by each producer; each producer needs to be known and trusted by the CAS server, but not by each consumer. Of course, the CAS server itself is a potential bottleneck and single point of failure, but standard replication techniques can be used to address this concern.

This structure also addresses flexibility and expressibility by allowing producer-community agreements and community policies to be expressed directly within the CAS server. Thus, for example, it is straightforward for a provider to agree to provide 10% of their resources to a community, and for the community to decide to provide 30% of its aggregate resources for one purpose.

Finally, by externalizing policy enforcement into a third party server, it is possible to set up specialized policy servers, representing sub-communities within a VO or completely different VOs.

In the rest of this section, we provide more details on these interactions between the CAS and the different parties.

### 3.1 CAS: Community Perspective

An individual representing a community can instantiate a CAS server by acquiring an identity certificate for the CAS server, doing some initial configuration, and running the CAS server software. That individual can then send requests to the CAS server to enroll users and resources into the server (thereby identifying them as members of the community) and to create policy information (e.g., to create groups of users or resources, or to grant access on a group of resources to a group of users).

Depending on a community's policies, a CAS server may have a single administrator who controls everything, or it may take a more distributed approach. For example, having administrators in geographically distinct areas who can enroll users but not add them to groups, and allowing Principal Investigators to maintain the membership of groups that represent people working on their projects but not enroll new users.

In our implementation, management of the CAS server is supported through a command-line tool (useful for automated processes) as well as a GUI interface. CAS also supports the notion of groups, both of users and resources, thus allowing the community to have different roles within the community and to grant members the rights of those roles by assigning them to the appropriate groups.

Community users request a capability for a set of actions from the CAS server using the CAS client library or tools built on it. The user uses the capability to authenticate to a resource server and exercise the rights described by the capability. The request action can be easily integrated with existing applications, as we discuss in Section 6, either by wrapping the application in a program call that makes the appropriate request or by adding to the application a call to a simple client API we have developed.

### 3.2 Resource Provider Perspective

A resource provider that wishes to accept credentials from CAS servers must be able to enforce not only its own local policies but also the community policies carried in CAS credentials. To accomplish this task we designed and implemented a policy evaluation API. Server software on the resource must be modified to use this API for parsing and evaluating the policy statements contained in CAS credentials.

Once a resource provider has installed the appropriate resource server software, that provider can grant access to specific resources to specific communities by using local access-control mechanisms to grant access to those resources to the subject names associated with those communities' CAS servers. Prior to granting that access, the resource providers may use offline methods to verify that the CAS server is run by someone who actually represents the community, and that the community's policies are compatible with those of the resource provider. For example, the resource provider may require that the CAS server administrators verify that new users agree to a certain Acceptable Use Policy before being enrolled into the CAS server.

## 4 Enabling Mechanisms

The development of a CAS implementation requires three major extensions to GSI mechanisms (Section 2): restricted proxy credentials that allow for fine-grained control of delegated rights; a policy language for specifying the rights carried in the restricted proxy credentials, and libraries and APIs to facilitate the delegation of restricted proxies by the CAS server and the enforcement of proxy restrictions by resource providers.

### 4.1 Restricted Proxy Credentials

The CAS server grants rights to community members by using GSI delegation mechanisms to grant them proxy credentials. GSI originally supported only a simple form of delegation, namely impersonation. However, in most cases it is inappropriate for the CAS server to delegate all of its authority to a user, because a community's access control policy usually grants different sets of rights to different users.

We have extended the GSI delegation feature to support rich restriction policies to allow grantors to place specific limits on rights that they grant. We accomplished this by defining extensions to X.509 Certificates to carry restriction policies [20]; we call a proxy carrying such a restriction policy a *restricted proxy*. CAS servers use these restricted proxies to delegate to each user only those rights granted to that user under the community's policy.

The CAS server uses restricted proxy credentials to delegate to each user only those rights granted to that user by the community policy. The CAS server delegates the user a restricted proxy credential that both authorizes the user to act as a member of the community and limits what the user can do as part of that community.

Some applications base authorization decisions on the comparison of two identities (for example, a peer-to-peer application may permit access if and only if the remote and local identities are the same). This comparison becomes meaningless for proxy credentials, because the same entity may grant proxies to several different individuals. We have implemented a hierarchical "proxy group" mechanism that enables the grantor of a proxy credential to associate a group name with each proxy certificate it grants, so that these applications can use the combination of issuer identity and proxy group (for example, a peer-to-peer application may permit access if and only if both the remote and local issuer identities and proxy groups are the same). The CAS server uses a different proxy group for each session.

### 4.2 Policy Language

Our restricted proxy credential format is designed to be neutral to the actual policy language employed and can support arbitrary policy languages such as Controlled English [21], ASL [22], or Ponder [23]. In our specification we do not state a specific language to be used, but instead have a field in restricted proxies that specifies the language of the policy carried by the proxy. GSI also treats the policy as opaque, meaning that only the creator of a policy and resources enforcing it need to understand it. This allows us to evolve our policy language over time as new requirements are understood, and as policy languages themselves evolve.

For our initial implementation of CAS, we chose to implement a simple policy language consisting of a list of the rights granted. Each right consists of a list of object names and a list of allowed actions on those objects. Although this simple language is obviously not rich enough to cover the whole range of applications, it has proven useful in our initial application case study (Section

6). We will continue to evaluate and compare various existing and emerging policy languages for their applicability to Grid applications.

### 4.3 Libraries and APIs

Policies carried in restricted proxy credentials need to be evaluated by resources accepting these credentials for authentication. To accomplish this task we designed and implemented a policy evaluation API and library. Internally our implementation uses the Generic Authorization and Access control (GAA) API [24] because of its ability to be configured to allow pluggable modules for acquiring, parsing and evaluation of policy; this flexibility is an essential requirement for supporting new policy languages that we may choose to use in the future.

GSI uses the Generic Security Services API (GSSAPI) [25] as its API, with extensions that we have designed and developed to support security features required in advanced Grid applications, specifically delegation flexibility and mechanisms to extract information. We have documented our extensions in an Global Grid Forum draft [26] to encourage their implementation in other GSSAPI libraries.

## 5 Implementation

We have implemented a CAS server, administrative clients for managing community policy information, and end-user client applications. These programs are all written in python and built on the pyGlobus wrappers [27] for the Globus Toolkit.

We have developed an Authorization API and library that services accepting CAS credentials can use to evaluate the policies contained in those credentials. We have also modified an FTP server to use this authorization library and accept restricted proxy credentials.

As we describe in the following section we have then built on these tools to integrate CAS into a real application.

## 6 Earth System Grid Case Study

We describe here an early CAS application, namely file access control within the Earth System Grid (ESG) [28], a distributed network of storage systems containing environmental data. In particular, we integrated CAS access control with the Visual Climate Data Analysis Tool (VCDAT) [29], an interactive tool that allows environmental scientists to select from and visualize a large collection of (potentially replicated) climate data.

ESG data is stored in a distributed system consisting of a metadata service, a replica manager, and a number of geographically and administratively distributed FTP servers. The metadata service lists available datasets, information about the data in each set, and the logical filenames of the datasets. The replica manager [30] maps from logical filenames to physical locations (i.e., hostnames and paths). Datasets are stored on the FTP servers, with each dataset generally replicated on multiple servers for locality and reliability. VCDAT processes user requests by consulting the metadata service to discover available datasets. It presents this information to the user who selects one or more datasets for visualization. For each dataset selected, VCDAT consults the replica manager, discovers where the data is located, selects an FTP server, downloads the data from the server, and then renders it for the user.

Prior to the work described here, access control within VDCAT was handled via manual updating of access control lists at individual FTP sites. As discussed earlier, this approach has significant scalability and usability difficulties. There are also expressiveness limitations: for example, some Unix systems cannot enforce the policy "these five people may read this file, and these three people may write it."

A CAS server solves these problems by giving both users and resource administrators a single point of contact for dealing with each other. For example, when a resource administrator decides to make resources available to the ESG community, they first grant access to the ESG CAS server using their existing local mechanisms. In the ESG case, this consists of creating a Unix account, using standard GSI mechanisms to map the ESG CAS server's subject name to that Unix account, and granting file permissions to that account. Note that the administrator needs to perform this process only once. If the resource administrator later decides to grant the community access to additional resources, or to revoke the community's access to some resources, then that administrator can do so using the same local mechanism (in this case, by changing file system permissions). The resource administrator does not need to be involved when individuals join or leave the community, or when an individual's role within the community changes.

A new ESG user needing access to the climate data now needs only to go to the ESG CAS administrator to obtain needed rights. The CAS administrator simply adds the user to the CAS database, putting them in the groups appropriate to the ESG community policy. The ESG CAS administrator can be someone more closely tied with the ESG community than the FTP administrators and hence more familiar with ESG users and the community policies. Depending on the community's policies, the CAS administrator may also delegate some of the responsibility for maintaining the CAS database—for example, the CAS administrator may grant the authority to enroll new users to several people at different geographic locations, or may grant the principal investigator of a project the authority to add people to or delete people from a group of project participants.

A modified VCDAT client contacts the ESG CAS server before downloading the data and retrieves a CAS credential granting rights to access the data. It then uses the CAS credential to authenticate to the FTP server and retrieve the data using standard Globus software. All this extra activity involving the CAS is performed transparently to the user.

We have prototyped the above CAS-enabled system and verified it works as expected. Modifying the VCDAT client was trivial, requiring only a dozen or so lines of code changes.

# 7 Security Considerations

## 7.1 Restricted Proxy Certificates

The security implications of restricted proxy certificates are discussed in detail in [20]. To summarize, we have tried to ensure that an entity cannot delegate more authority than it has and that a server process that does not know how to enforce the restrictions in a restricted proxy certificate will reject the certificate outright. The effective validity time for a proxy certificate (restricted or otherwise) is the intersection of the validity times of all the certificates in the certificate chain; the effective set of allowed operations is the intersection of what's allowed by all the certificates in the chain. Proxy restrictions are encoded in a critical X.509 extension, so restricted proxies are rejected by authentication libraries that don't understand restrictions. The authentication libraries that do understand restrictions reject restricted proxies unless the calling program has indicated its willingness to enforce proxy restrictions.

We do not currently provide a mechanism for the revocation of proxy certificates, relying instead on their short lifetimes.

## 7.2 Compromised CAS server

If a CAS server is compromised or untrustworthy, it can issue credentials that do not reflect the policies of the community that it represents; for example, it may grant access to people who are not members of the community. A resource provider would honor such a credential for any access that its local policies grant to that community.

A compromised CAS server might also issue credentials that (attempt to) grant access to resources that don't belong to the community, but unless a resource server has been configured to grant access on those resources to that community, those credentials will be rejected.

If a CAS server is discovered to have been compromised, resource servers can use their local access control mechanisms to revoke any permissions granted to that server.

## 7.3 Revocation Mechanisms

If a user credential is compromised, then that user can be unenrolled from the CAS server (i.e., removed from the list of users in the community). The CAS server will then refuse to delegate any credentials to that user; however, any credentials previously delegated to that user will continue to be honored until they expire.

Credentials issued by the CAS server to community members, like most GSI proxy credentials, are given relatively short life spans, usually on the order of hours. This allows these credentials to be used without the use of a revocation mechanism since they generally expire rapidly enough for most applications.

## 7.4 Compromised Resource Server

Although a compromised or untrustworthy resource server is likely to be a serious problem (e.g., if a community stores sensitive files on that server), this does not create cascaded security issues. For example, if a user uses a CAS credential to authenticate to a compromised resource server, that server cannot use that CAS credential to gain additional access, because the resource server never sees the private key.

For highly-sensitive applications where greater assurance of resource enforcement of community policy is required, a mechanism such as Law-Governed Interaction [31] can be used to help assure this.

# 8 Related Work

The Akenti system [7, 8] identifies a set of *stakeholders* with a resource, where each stakeholder is allowed to place restrictions on who and how the resource can be used. These restrictions are specified in terms of what attributes a user must possess in order to perform specific requests. If all stakeholders approve a request, then the request may be performed. Akenti makes extensive use of PKI certificates for encoding both user attributes as well as usage conditions. Our work is distinguished by its focus on supporting the centralized specification of community policies governing collections of resources, such as who is allowed to read and write replicated data in a Data Grid; Akenti, on the other hand, is concerned primarily with expressing the use conditions that govern access to individual resources. This different focus leads us to adopt different technical approaches. For example, in Akenti every resource must know about and trust the CA of every potential user, which seems to us to be a significant obstacle to scalability. In contrast, a CAS server must know about and trust the CA of every user, but individual resources need know about only the CAS server's CA. Similarly, a CAS server provides a centralized location at which can be collected the various use conditions that govern access to a resource; once these are verified, the resource need deal only with a single capability, rather than consulting a potentially large set of repositories to find all the applicable use conditions and attribute certificates. Finally CAS provides a mechanism to delegate permissions on a set of resources distributed across different administrative domains. We believe that for these reasons CAS represents an interesting alternative—and most likely complementary—technology to Akenti; for example, Akenti could be used as the "local" access mechanism in the CAS model.

The Secure Virtual Enclaves (SVE) infrastructure [32] provides a mechanism for the controlled sharing of resources among organizations. An enclave is a set of resources managed by the same organization. An SVE is a collection of subsets of one or more enclaves. Local enclave administrators grant permissions on *types* (groups of resources, or of operations on resources, within the local enclave) to *domains* (groups of users), possibly subject to additional conditions. The mapping of resources to types, and the granting of permissions on each type, is done within each local enclave and not shared with other enclaves in the SVE; however, the specification of domains is shared across enclaves within an SVE (i.e., if the administrator of any enclave within an SVE adds a user to a domain, that user will be considered part of that domain within the entire SVE). SVE provides "interceptors," middleware components for some software architectures, that enable the use of SVE without modifying application-level code.

The SVE approach has some features in common with the CAS approach. Both frameworks allow local resource administrators to determine the upper bound of what access is granted to the community, while allowing people other than the local administrators to specify some parts of the effective policy. Both frameworks also support the aggregation of users and resources. The SVE approach allows an enclave to export resources to more than one SVE; the CAS approach allows resource providers to grant permissions to more than one community. However, there are significant differences. The SVE model seems to be focused on relationships among organizations, while the CAS model is focused on the relationships between organizations and communities and among individuals within a community. In the SVE model, all policy is managed by enclave administrators, and all enclave administrators have equal authority to maintain the membership of domains (user groups). In the CAS model, course-grained access control is maintained by local resource administrators, and fine-grained access control is maintained by a community representative, who may delegate out subsets of that authority; neither the original community representative nor anyone to whom this authority is delegated is required to be a resource administrator. This delegation of administrative authority may also be very fine-grained; for example, a community administrator may create a group and delegate to an individual the authority to add established community members to and delete them from that group, without giving that person any other administrative authority. The SVE model allows aggregation of resources, but only within an enclave; the CAS model does not have that restriction. We have found that communities often wish to grant permissions to collections of resources that belong to different organizations; for example, in data grids [33], multiple replicas of a file, which generally should have identical access control policies, typically exist on servers in different organizations. The SVE project has concentrated more than the CAS project on programming frameworks.

The theoretical concepts of proxy certificates, restricted proxies, and authorization servers that generate proxy certificates were described in [10].

## 9 Future Directions

### 9.1 Accounting

Many policies contain quotas or similar rights that are dynamic in that they depend on a user's current resource consumption. For example a user may be assigned a quota on the total amount of storage they can consume by community policy, meaning the amount of new data a user can store is the difference between the quota and amount they have currently stored.

CAS currently works with statically defined rights. For CAS to enforce these policies a distributed accounting system must be put into place to provide feedback on the user's resource consumption across all resource servers. The development of such a system can be expected to be complicated, due to the need to communicate usage information from resources to CAS as well for standard methods for describing resource usage.

We plan to experiment with an approach to this problem based on tagging each restricted proxy issued by the CAS with a Globally Unique ID (GUID). This GUID will be used by resources to track resource consumption and report it back to CAS. CAS will then map the GUID back to the original user and keep track of the resources being consumed by the user.

Additional areas we intend to research that are needed to complete the accounting system are tools for allowing resources to log usage, protocols and tools for distributing this accounting information back to the CAS, and policy languages for expressing limits on the amount of resource consumption allowed by a user.

### 9.2 Delegation tracing

While thinking of an individual as simply a member of a community is acceptable for the purpose of authorization, there are instances where resources will want to know the actual identity of a user. Auditing is a common reason for this, so that if malicious behavior is attempted it can be tracked back to an individual.

To enable this functionality we are investigating methods of tracing delegations. This would allow a party accepting a credential to be able to determine the identity of the party to whom the credential was delegated and hence the identity of the user who accepted the credential from CAS.

## 9.3 Replication of CAS Server

A system based on a single CAS server for a community may have the problem of being a single point of failure as well as a potential performance bottleneck. We will explore methods of replicating the CAS server to alleviate this potential problem. The best approach will depend on how often we see the community policy changed in practice. If community policy tends to be changed infrequently we can define a single master server that can accept changes and then routinely replicates the policy to one or more read-only slave servers. If the community policy changes frequently we will require a more complicated distribution among a set of peer servers where all can act to update the policy and the loss of any one server does not lead to a loss of functionality.

## 10 Summary

This paper describes the Community Authorization Service (CAS) we have developed to solve three critical authorization problems that arise in distributed virtual organizations: scalability, flexibility and expressibility, and the need for policy hierarchies. We address these problems by introducing a trusted third party administered by the virtual organization that performs fine-grain control of community policy while leaving ultimate control of resource access with the resource owners. We also describe our experience integrating CAS with a real world application. This experience provides us with some initial evidence that CAS is a viable solution to our target problems.

## 11 Acknowledgments

We thank Doug Engert, Sam Meder, Chris Nebergall, and Shubi Raghunathan for contributions to the CAS project. Apologies to anyone we missed. We also thank the anonymous reviewers for their insightful critiques.

This work was supported in part by the Mathematical, Information, and Computational Sciences Division subprogram of the Office of Advanced Scientific Computing Research, Office of Science, SciDAC Program, U.S. Department of Energy, under contracts W-31-109-Eng-38.